\documentclass[conference]{IEEEtran}

\usepackage{cite}
\usepackage{amsmath,amssymb,amsfonts}
\usepackage{graphicx}
\usepackage{xcolor}
\usepackage{array,multirow}
\usepackage{tablefootnote}
\usepackage{threeparttable}
\usepackage{makecell}
\usepackage{booktabs, makecell, tabularx}
\usepackage{soul}

\def\BibTeX{{\rm B\kern-.05em{\sc i\kern-.025em b}\kern-.08em
    T\kern-.1667em\lower.7ex\hbox{E}\kern-.125emX}}

\makeatletter
\IEEEtriggercmd{\reset@font\normalfont\fontsize{7.9pt}{8.60pt}\selectfont}
\newcommand*\titleheader[1]{\gdef\@titleheader{#1}}
\AtBeginDocument{%
  \let\st@red@title\@title
  \def\@title{%
    \bgroup\normalfont\normalsize\centering\@titleheader\par\egroup
    \vskip1ex\st@red@title}
}
\makeatother
\IEEEtriggeratref{1}

\titleheader{\vspace{-30pt}To appear at the 31st IEEE Int'l. Conference on Electronics Circuits and Systems (ICECS), 18-20 Nov, 2024, Nancy, France.}
\title{MPAI: A Co-Processing Architecture 
with MPSoC \& AI Accelerators for Vision Applications in Space}%

\makeatletter
\def\ps@IEEEtitlepagestyle{
  \def\@oddfoot{\mycopyrightnotice}
  \def\@evenfoot{}
}
\def\mycopyrightnotice{
  {\footnotesize
  \begin{minipage}{\textwidth}
  \centering%
  ~\copyright~2024 IEEE.  Personal use of this material is permitted.  Permission from IEEE must be obtained for all other uses, in any current or future media, including reprinting/republishing this material for advertising or promotional purposes, creating new collective works, for resale or redistribution\\to servers or lists, or reuse of any copyrighted component of this work in other works.
  \end{minipage}
  }
}

\begin{document}

\author{\IEEEauthorblockN{%
Vasileios Leon\IEEEauthorrefmark{1},
Panagiotis Minaidis\IEEEauthorrefmark{1},  
Dimitrios Soudris\IEEEauthorrefmark{1}, 
George Lentaris\IEEEauthorrefmark{2}\IEEEauthorrefmark{1}}
\vspace{2pt}
\IEEEauthorblockA{\IEEEauthorrefmark{1}\emph{National Technical University of Athens, School of Electrical and Computer Engineering, Zografou 15780, Greece}}
\vspace{0.7pt}
\IEEEauthorblockA{\IEEEauthorrefmark{2}\emph{University of West Attica, Department of Informatics and Computer Engineering, Egaleo 12243, Greece}}
\vspace{2pt}
\IEEEauthorblockA{%
\fontsize{9.5}{10}\selectfont
Emails:
\IEEEauthorrefmark{1}\{vleon, pminaidis, dsoudris\}@microlab.ntua.gr,
\IEEEauthorrefmark{2}glentaris@uniwa.gr}}
    
\maketitle

\begin{abstract}
The emerging need for fast and power-efficient AI/ML deployment on-board spacecraft has forced the space industry to examine specialized accelerators, which have been successfully used in terrestrial applications. Towards this direction, the current work introduces a very heterogeneous co-processing architecture that is built around UltraScale+ MPSoC and its programmable DPU, as well as commercial AI/ML accelerators such as MyriadX VPU and Edge TPU. The proposed architecture, called \emph{MPAI}, handles networks of different size/complexity and accommodates speed--accuracy--energy trade-offs by exploiting the diversity of accelerators in precision and computational power. This brief provides technical background and reports preliminary experimental results and outcomes. 
\end{abstract}

{\linespread{0.995}
}

\section{Introduction}
The advent of the NewSpace era has been accompanied by the emergence of novel space applications,
which are based on demanding AI/ML and DSP computations. 
As a result, 
traditional computing paradigms and architectures 
are stressed to meet the high-performance requirements
in applications,
among others, 
of Earth observation \cite{fsat_myriad}, vision-based navigation \cite{panous}, and satellite communications \cite{camad2024}.
To reach the performance goals, 
heterogeneous co-processing architectures are being examined,
which also rely on Commercial Off-The-Shelf (COTS) devices \cite{cots_space}
for the workload acceleration.  
At the same time, such architectures tend to meet the requirements for low power, enhanced adaptability to mission scenarios, and improved in-flight re-programmamability. 

For many years, the space industry has been using FPGAs for on-board payload data processing \cite{access_brave}.
The recent development of specialized hardware accelerators
has brought AI to the fore for space. 
As a result,  
there is a plethora of works
evaluating the suitability
of AI accelerators, such as Edge TPU \cite{edhpc} and MyriadX VPU \cite{leon_elsi}, to be on-board payload data processors  (e.g., with respect to performance, power, cost, and radiation resilience).  
This radical shift for on-board  processors
is already apparent;
spaceborne computing platforms with AI accelerators 
have been developed (e.g., Ubotica's CogniSAT-XE2 with MyriadX VPU and NASA's SpaceCube with Edge TPU),
while there are active satellite missions
with on-board AI accelerators
(e.g., $\mathrm{\Phi}$-Sat-1 with Myriad2 VPU\cite{fsat_myriad}).

In this modified landscape of on-board processing in space,
the current work proposes \emph{MPAI}
(\emph{MP}SoC\hspace{2pt}+\hspace{2pt}\emph{AI}),
which is 
a COTS co-processing architecture based on 
the UltraScale+ MPSoC FPGA
and an AI/ML accelerator 
(e.g., MyriadX VPU or Edge TPU). 
This heterogeneous architecture can serve a great variety
of workloads and tasks: 
instrument/sensor handling (MPSoC),
general-purpose computing (MPSoC's ARM CPUs),
acceleration of classic DSP functions (MPSoC's PL),
and
acceleration of AI/ML computations (MPSoC's PL-based DPU and/or VPU/TPU).  
The goal of this brief 
is to present the concept of MPAI,
discuss key technical issues,
and
report preliminary comparative results.
The evaluation focuses on demanding DNNs
for computer vision applications in space. 
However, 
MPAI can be also used in terrestrial applications
and for different types of AI networks  
(provided that the accelerators support their operations). 

\section{MPSoC-\&-AI Co-Processing Architecture}

\begin{figure}[b]
\vspace{-8pt}
    \centering
    \includegraphics[width=\columnwidth]{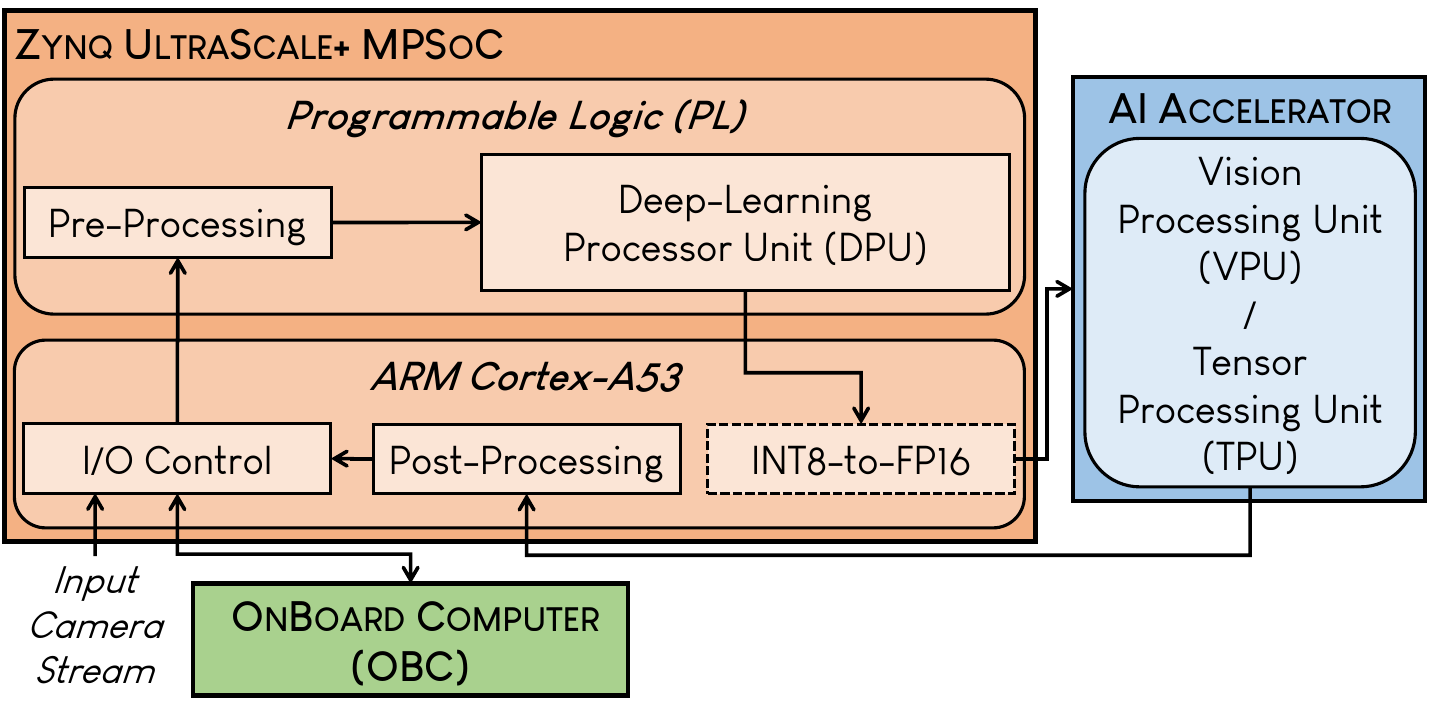}
    \vspace*{-22pt}
    \caption{Co-processing architecture with MPSoC and VPU/TPU AI accelerator.}
    \label{fige}
\end{figure}

The MPAI architecture,
illustrated in Fig. \ref{fige}, 
is based on MPSoC (CPU\hspace{1pt}+\hspace{1pt}PL\hspace{1pt}+\hspace{1pt}DPU) and an AI/ML accelerator (VPU, TPU, or other).
MPSoC receives the camera input to be processed  
and handles the communication with the on-board computer. 

The DPU is a softcore IP of AMD/Xilinx
that implements a programmable engine in PL  
for inferencing DNNs. 
The design is based on a deep pipelined 8-bit (INT8) architecture, 
with the 
processing elements taking full advantage of the fine-grained building blocks
(e.g., multipliers and accumulators). 
The on-chip memory is used for storing  
input activations, intermediate feature-maps, and output meta-data.
Data reuse is applied to reduce external memory bandwidth requirements. 
An instruction scheduler fetches instructions from the off-chip memory to control the operation of the engine. 
The instructions are generated by the Vitis AI compiler, which performs 
optimizations (e.g., layer fusion) in the network graph and generates the executable code.

The MyriadX VPU offers a great heterogeneity in processors, memories and I/O interfaces. It integrates 2 general-purpose LEON4 CPUs, 16 SIMD \& VLIW programmable cores, various hardware imaging filters, and a dedicated AI accelerator engine. The memory hierarchy includes on-chip DDR DRAM, scratchpad SRAM, and caches. Specifically for AI/ML inference, the OpenVINO toolkit is used for fast deployment on the AI engine, performing model optimizations on the network’s frozen graph (e.g., network pruning, linear operation fusing, and grouped convolution fusing). The models are built on 16-bit floating-point (FP16) arithmetic. MyriadX comes in two variants: SoC and USB. 

The Edge TPU relies on a systolic array of multipliers \& accumulators for DNN acceleration, and an on-chip SRAM for storing the model's parameters and executable. There is a USB variant that integrates only the TPU accelerator, as well as various SoM variants that add ARM processors, DDR \& flash memories, and I/O interfaces. The models are first quantized to 8-bit integer (INT8)  using the TensorFlow Lite toolflow, and then they are compiled with the Edge TPU compiler. 

\section{Preliminary Experimental Results}

For preliminary evaluation,  
the following devices are used:
(i) the ZCU104 board featuring MPSoC and implementing two instances of the DPUCZDX8G DPU,
(ii) the Coral DevBoard single-board computer featuring the Edge TPU SoM,
and 
(iii) the NCS2 USB accelerator featuring the MyriadX VPU. 

\begin{figure}[!t]
\vspace{-2pt}
    \centering
    \includegraphics[width=0.92\columnwidth]{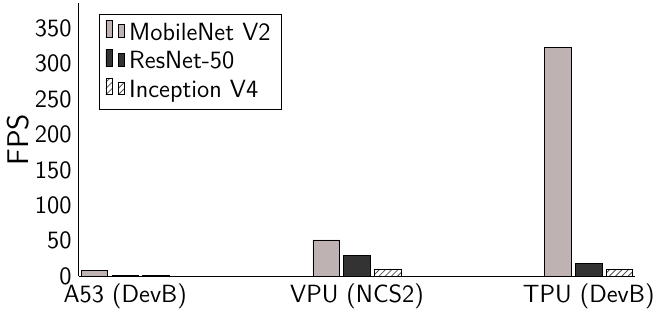}
    \vspace*{-11pt}
    \caption{Inference throughput of AI accelerators.}
    \label{fige2}
    \vspace{-6pt}
\end{figure}

The generic performance of the AI accelerators is illustrated in 
Fig. \ref{fige2} 
for three networks of different complexity/size.
For small networks (MobileNet V2), 
TPU provides 
8$\times$ more Frames Per Second (FPS) than VPU.
However, for a larger network (ResNet-50),
VPU 
delivers 2$\times$ throughput,
while for Inception
V4, both accelerators sustain $\sim$10 FPS.

Table \ref{tbe} reports results
from the 
the acceleration of the compute-intensive UrsoNet DNN \cite{ursonet},
which performs 
satellite pose estimation on the ``soyuz\_easy'' dataset. 
The individual results of each accelerator
include pre-processing tasks (e.g., image resampling) and DNN inference. 
The DPU delivers 3.8$\times$ and 2.8$\times$ speedup versus the VPU and the TPU, respectively. 
However, it is worse in terms of accuracy, as 
the LOCE and ORIE metrics are increased (even though TPU also uses INT8 precision). 
The MPAI approach (DPU\hspace{1pt}+\hspace{1pt}VPU) is configured using partition-aware model training:
the convolutional layers are executed on the DPU with INT8,
while the fully connected layers that calculate the satellite location and orientation
are executed on the VPU with FP16.
Namely, the demanding heavyweight layers are accelerated with the fastest DPU,
while the fully-connected layers, which significantly affect the accuracy, 
are executed on the VPU with better precision. 
The MPAI latency is 2.7$\times$ and 2$\times$ 
better than that of the full execution on the VPU and TPU, respectively. 
Compared to the DPU, 
it is slightly worse, but the MPAI accuracy almost matches the baseline model accuracy. 

\section{Conclusion \& Future Work}
In this brief, an heterogeneous, mixed-precision co-processing architecture for accelerating AI/ML in space applications was presented. 
The advantage of this architecture is twofold:
(i) it can be implemented in a single computing board integrating CPU\hspace{1pt}+\hspace{1pt}PL\hspace{1pt}+\hspace{1pt}DPU\hspace{1pt}+\hspace{1pt}VPU/TPU,
and  
(ii) it efficiently accommodates various scenarios and complies with different system requirements for speed, accuracy, and energy consumption.  
Future work will focus on deploying virtualization \& orchestration functionalities, 
extracting a methodology and design guidelines for the model partitioning and accelerator selection,
and evaluating additional AI/ML workloads.

\begin{table}[!t]
\renewcommand{\arraystretch}{1.05}
\setlength{\tabcolsep}{2.95pt}
\caption{Benchmarking Results for Satellite Pose Estimation DNN\\on 1280$\times$960$\times$3 Images}
\vspace{-5pt}
\label{tbe}
\hspace{-5pt}
\begin{threeparttable}
\begin{tabular}{ccccccc}
\hline 
\multirow{2.2}{*}{\textbf{Processor /}} & \multirow{2.2}{*}{\textbf{Hosting}} & \multirow{2.2}{*}{\textbf{Model}} & \multicolumn{2}{c}{\textbf{Accuracy}\tnote{\tiny1}} & \multicolumn{2}{c}{\textbf{Latency}} \\
\cmidrule(lr){4-5}\cmidrule(lr){6-7}
\textbf{Accelerator} & \textbf{Device} & \textbf{Precision} & \emph{LOCE} & \emph{ORIE} & \emph{Inference} & \emph{Total} \\ 
\hline 
\hline 
Cortex-A53 CPU  & DevBoard     & FP32 & 0.68\hspace{1pt}m & 7.28$^\circ$ & 9890\hspace{1pt}ms  & 9928\hspace{1pt}ms \\ 
Cortex-A53 CPU    & ZCU104      &  FP16 & 0.87\hspace{1pt}m & 8.09$^\circ$ & 4210\hspace{1pt}ms & 4338\hspace{1pt}ms \\ 
MyriadX VPU       & NCS2        &  FP16 & 0.69\hspace{1pt}m & 8.71$^\circ$ & 246\hspace{1pt}ms  & 252\hspace{1pt}ms \\ 
Edge TPU           & DevBoard   &  INT8 & 0.66\hspace{1pt}m & 7.60$^\circ$ & 149\hspace{1pt}ms    & 187\hspace{1pt}ms \\ 
MPSoC DPU           & ZCU104     &  INT8& 0.96\hspace{1pt}m & 9.29$^\circ$ & 53\hspace{1pt}ms     & 66\hspace{1pt}ms \\ 
\hline 
\multirow{2}{*}{DPU\hspace{1pt}+\hspace{1pt}VPU} & ZCU104    & INT8 & \multirow{2}{*}{0.68\hspace{1pt}m} & \multirow{2}{*}{7.32$^\circ$} & \multirow{2}{*}{79\hspace{1pt}ms} & \multirow{2}{*}{92\hspace{1pt}ms} \\[-1.5pt] 
 & +\hspace{1pt}NCS2  & +\hspace{1pt}FP16 &  &  &  &  \\ 
\hline
\end{tabular}
 \begin{tablenotes}[flushleft]
   \item[\tiny1]{\fontsize{6.3}{7.7}\selectfont 
   \underline{Baseline SW Algorithm}:
   $\cdot$ Localization Error (LOCE) = 0.63\hspace{1pt}m
   $\cdot$ Orientation Error (ORIE) = 7.20$^\circ$}
   \end{tablenotes}
  \end{threeparttable}
 \vspace{-7pt}
\end{table}
 
\bibliographystyle{IEEEtran}
\bibliography{REFERENCES.bib}

\end{document}